\documentclass[12pt]{article}
\usepackage{axodraw,bbold}

\catcode`@=12
\begin{document}
\thispagestyle{empty}
\begin{flushright} 
UCRHEP-T450\\ 
April 2008\
\end{flushright}
\vspace{0.5in}
\begin{center}
{\LARGE	\bf Common Origin of (--)$^{\bf L}$, (--)$^{\bf 3B}$, and\\ Strong CP 
Conservation\\}
\vspace{1.0in}
{\bf Ernest Ma\\}
\vspace{0.2in}
{\sl Department of Physics and Astronomy, University of California,\\}
\vspace{0.1in}
{\sl Riverside, California 92521, USA\\}
\vspace{1.5in}
\end{center}

\begin{abstract}\
The multiplicative conservation of both lepton and baryon numbers, i.e. 
$(-)^{L}$ and $(-)^{3B}$, is connected to an axionic solution of the strong 
CP problem in a supersymmetric, unifiable model of quark and lepton 
interactions.  New particles are predicted at the TeV scale, with 
verifiable consequences at the Large Hadron Collider.
\end{abstract}

\newpage
\baselineskip 24pt

Experimentally, there is no evidence against the conservation of additive 
lepton ($L$) and baryon ($B$) numbers.  Nevertheless, the prevailing 
theoretical thinking is that neutrino masses are Majorana and only 
$(-)^L$ is conserved.  This implies the occurrence of neutrinoless double 
beta decay \cite{ev02} which is being pursued actively but yet to be 
confirmed. Recently it has been pointed out \cite{m08} that the parallel 
situation of $(-)^{3B}$ conservation is also possible, with the consequence 
of an absolutely stable proton but allowing deuteron decay and 
neutron-antineutron oscillations.  In the following these multiplicative 
conservation laws are connected to an axionic solution of the strong CP 
problem in a supersymmetric, unifiable model of quark and lepton 
interactions. Heavy quarks of charge $\mp 1/3$ with $B = \mp 2/3$ at 
the TeV scale are predicted.

The idea of $(-)^L$ conservation is well-known.  The neutrino $\nu$ $(L=1)$ 
is paired with a singlet neutral fermion $N^c$ $(L=-1)$ through the standard 
Higgs doublet $(\phi^+,\phi^0)$.  In the presence of electroweak $SU(2)_L 
\times U(1)_Y$ symmetry breaking, $\langle \phi^0 \rangle = v$ implies a 
Dirac mass $m_D$ linking $\nu$ with $N^c$.  However, $N^c$ is a gauge singlet 
and as such, it is allowed a large Majorana mass $m_N$, thereby breaking 
$L$ to $(-)^L$ and resulting in a small seesaw Majorana mass for $\nu$, 
i.e. $m_\nu = m_D^2/m_N$.

Similarly, the idea of $(-)^{3B}$ conservation requires a singlet neutral 
fermion $\Sigma$, carrying $B=1$ in the effective interaction
$$u^c_i d^c_j d^c_k \Sigma.$$
To implement this in a renormalizable theory, the simplest way is to 
introduce singlet scalar quark fields $\tilde{h}, \tilde{h}^c$ with 
charges $\mp 1/3$ and $B=\mp 2/3$, so that the interactions $u^c d^c 
\tilde{h}^c$ and $\tilde{h} d^c \Sigma$ are allowed.  Alternatively, 
$\tilde{h}, \tilde{h}^c$ may be assigned charges $\pm 2/3$, in which case 
$d^c d^c \tilde{h}^c$ and $\tilde{h} u^c \Sigma$ are allowed.  The large 
Majorana mass $m_\Sigma$ breaks $B$ to $(-)^{3B}$ under which the usual 
quarks are odd and the exotic scalar quarks $\tilde{h}, \tilde{h}^c$ are 
even.

The decay of the lightest $N^c$ in the early Universe generates a lepton 
asymmetry, whereas the decay of the lightest $\Sigma$ generates a baryon 
asymmetry.  Both are converted to a $B-L$ asymmetry from the intervention 
of electroweak sphalerons \cite{dnn08}.  Below the scale of $m_N$, all 
particle interactions conserve additive $L$ except for rare processes 
involving the effective exchange of $N^c$ such as neutrinoless double beta 
decay.  Below the scale of $m_\Sigma$, all particle interactions conserve 
additive $B$ except for rare processes involving the effective exchange of 
$\Sigma$ such as deuteron decay and neutron-antineutron oscillations.  The 
two scales $m_N$ and $m_\Sigma$ are not {\it a priori} related, but in the 
context of an axionic solution of the strong CP problem, both will come 
from the vacuum expectation value of a singlet field, the dynamical phase 
of which contains the axion, as shown below.  A unifying picture is thus 
possible for the common origin of $(-)^L$, $(-)^{3B}$, and strong CP 
conservation.

The strong CP problem is the appearance of the instanton-induced term 
\cite{cdg76,jr76}
\begin{equation}
{\cal L}_\theta = \theta_{QCD} {g_s^2 \over 64 \pi^2} \epsilon_{\mu \nu \alpha 
\beta} G_a^{\mu \nu} G_a^{\alpha \beta}
\end{equation}
in the effective Lagrangian of quantum chromodynamics (QCD), where $g_s$ is 
the strong coupling constant, and
\begin{equation}
G_a^{\mu \nu} = \partial^\mu G_a^\nu - \partial^\nu G_a^\mu + g_s f_{abc} G_b^\mu 
G_c^\nu
\end{equation}
is the gluonic field strength.  This term is odd under CP and if 
$\theta_{QCD}$ is of order unity, the neutron electric dipole moment would be 
$10^{10}$ times its present experimental upper limit ($0.63 \times 10^{-25} e$ 
cm) \cite{h99}.  This undesirable situation is most elegantly resolved by 
invoking a dynamical mechanism \cite{pq77} to relax the above $\theta_{QCD}$  
parameter (including all contributions from colored fermions) to zero.  
However, this requires an anomalous global $U(1)_{PQ}$ symmetry which is 
broken at the scale $f_a$ and results necessarily \cite{we78,wi78} in a 
very light pseudoscalar particle called the axion, which has not yet been 
observed \cite{rb00}.

To reconcile the nonobservation of an axion in present experiments and the 
constraint $10^9$ GeV $< f_a < 10^{12}$ GeV from astrophysics and cosmology 
\cite{r99}, three types of ``invisible'' axions have been discussed. (I) 
The DFSZ solution \cite{dfs81,z80} introduces a heavy singlet scalar field 
as the source of the axion but its mixing with the doublet scalar fields 
(which couple to the usual quarks) is very much suppressed. (II) The KSVZ 
solution \cite{k79,svz80} also has a heavy singlet scalar field but it 
couples only to new heavy colored fermions. (III) The gluino solution 
\cite{dm00} identifies the $U(1)_R$ of superfield transformations with 
$U(1)_{PQ}$ so that the axion is a dynamical phase attached to the gluino 
(which contributes to $\theta_{QCD}$ because it is a colored fermion) 
as well as all other superparticles.  

In a supersymmetric extension of the Standard Model, it is also important 
that the breaking of $U(1)_{PQ}$ at the large scale $f_a$ does not break 
supersymmetry as well.  This may be accomplished using three singlet 
superfields in various ways, for the gluino solution \cite{dms00,dm01,m07} 
and for the DFSZ solution \cite{m01}.  In Table 1, the $(-)^L$, $(-)^{3B}$, 
and PQ charges of the superfields of this construction are listed.

\begin{table}[htb]
\caption{Particle content of proposed model.}
\begin{center}
\begin{tabular}{|c|c|c|c|c|}
\hline 
Superfield & $SU(3)_C \times SU(2)_L \times U(1)_Y$ & $(-)^L$ & $(-)^{3B}$ 
& $U(1)_{PQ}$ \\ 
\hline
$Q \equiv (u,d)$ & $(3,2,1/6)$ & + & -- & 1/2 \\ 
$u^c$ & $(3^*,1,-2/3)$ & + & -- & 1/2 \\ 
$d^c$ & $(3^*,1,1/3)$ & + & -- & 1/2 \\ 
$\Sigma$ & $(1,1,0)$ & + & -- & 1/2 \\
\hline
$L \equiv (\nu,e)$ & $(1,2,-1/2)$ & -- & + & 1/2 \\ 
$e^c$ & $(1,1,1)$ & -- & + & 1/2 \\ 
$N^c$ & $(1,1,0)$ & -- & + & 1/2 \\
\hline
$\Phi_1 \equiv (\phi^0_1,\phi^-_1)$ & $(1,2,-1/2)$ & + & + & --1 \\ 
$\Phi_2 \equiv (\phi^+_2,\phi^0_2)$ & $(1,2,1/2)$ & + & + & --1 \\ 
\hline
$h$ & $(3,1,-1/3)$ & + & + & --1 \\ 
$h^c$ & $(3^*,1,1/3)$ & + & + & --1 \\ 
\hline
$S_2$ & $(1,1,0)$ & + & + & 2 \\ 
$S_1$ & $(1,1,0)$ & + & + & --1 \\ 
$S_0$ & $(1,1,0)$ & + & + & --2 \\ 
\hline
\end{tabular}
\end{center}
\end{table}

The most general superpotential with this particle content is then given by
\begin{eqnarray}
W &=& m_0 S_0 S_2 + \lambda_1 S_1 S_1 S_2 + \lambda_2 S_1 N^c N^c + 
\lambda_3 S_1 \Sigma \Sigma \nonumber \\ 
&+& f_1 S_2 \Phi_1 \Phi_2 + f_2 S_2 h h^c + f_3 Q Q h + f_4 u^c d^c h^c + 
f_5 h d^c \Sigma \nonumber \\ 
&+& f_d \Phi_1 Q d^c + f_u \Phi_2 Q u^c + f_e \Phi_1 L e^c + f_N \Phi_2 L N^c.
\end{eqnarray}
Note that the only allowed mass term is $m_0$ which is thus expected to be 
large.  With $W$ of Eq.~(3), it has been shown \cite{m01} that it is 
possible to break $U(1)_{PQ}$ spontaneously at the scale $m_0$ without 
breaking the supersymmetry.  The soft breaking of supersymmetry will then 
introduce another (much smaller) scale $M_{SUSY}$, with the result 
$u_1 = \langle S_1 \rangle$ and $u_0 = \langle S_0 \rangle$ are of order 
$m_0$, whereas $u_2 = \langle S_2 \rangle$ is of order $M_{SUSY}$.  This 
means that the so-called $\mu$ problem in the Minimal Supersymmetric 
Standard Model (MSSM) is solved because $\mu = f_1 u_2$.  Similarly, the 
exotic $h$ quark has the mass $f_2 u_2$ and should be observable at 
the Large Hadron Collider (LHC).  As for the masses of $N^c$ and $\Sigma$, 
they are given by $2 \lambda_2 u_1$ and $2 \lambda_3 u_1$ respectively, 
with the axion contained in the dynamical phase of $S_1$.  Hence a common 
origin emerges for the conservation of $(-)^L$, $(-)^{3B}$, and strong CP.

To see how supersymmetry remains unbroken at the axion scale, consider 
the scalar potential of $S_{2,1,0}$, i.e.
\begin{equation}
V = m_0^2 |S_2|^2 + 4 \lambda_1^2 |S_1|^2 |S_2|^2 + 
|m_0 S_0 + \lambda_1 S_1^2|^2.
\end{equation}
There are two supersymmetric minima: the trivial one with 
$u_0 = u_1 = u_2 = 0$, and the much more interesting one with
\begin{equation}
u_2=0, ~~~ m_0 u_0 + \lambda_1 u_1^2 = 0.
\end{equation}
The latter breaks $U(1)_{PQ}$ spontaneously and shifting the superfields 
by $u_{2,1,0}$, the superpotential of $S_{2,1,0}$ becomes
\begin{equation}
W' = {m_0 \over u_1} (u_1 S_0 - 2 u_0 S_1) S_2 + \lambda_1 S_1 S_1 S_2,
\end{equation}
showing clearly that the linear combination
\begin{equation}
\chi = {u_1 S_1 + 2 u_0 S_0 \over \sqrt{u_1^2+4u_0^2}}
\end{equation}
is a massless superfield.

At this point, the individual values of $u_1$ and $u_0$ are not determined. 
This is because the vacuum is invariant not only under a phase rotation 
but also under a scale transformation as a result of the unbroken 
supersymmetry \cite{m99}, i.e. a flat direction.  As such, it is unstable 
and the soft breaking of supersymmetry at $M_{SUSY}$ will determine $u_1$ 
and $u_0$ separately, and $u_2$ will become nonzero.  Specifically, the 
supersymmetry of this theory is broken by all possible holomorphic soft 
terms which are invariant under $U(1)_{PQ}$.  As a result \cite{m01},
\begin{equation}
u_2 \sim M_{SUSY}, ~~~ m_0 u_0 + \lambda_1 u_1^2 \sim M^2_{SUSY},
\end{equation}
with $u_0$ and $u_1$ individually of order $m_0$.

As the electroweak $SU(2)_L \times U(1)_Y$ gauge symmetry is broken by the 
vacuum expectation values $v_{1,2}$ of $\phi^0_{1,2}$, the observed doublet 
neutrinos acquire naturally small Majorana masses given by $m_\nu = 
f_N^2 v_2^2/(2\lambda_2 u_1)$ by way of the usual seesaw mechanism. 
Since $\phi^0_{1,2}$ have PQ charges as well, the axion field is now 
given by
\begin{equation}
a = V^{-1} \left[ u_1 \theta_1 + 2 u_0 \theta_0 - 2 u_2 \theta_2 + 
{2 v_1 v_2 \over v_1^2+v_2^2} (v_1 \varphi_2 + v_2 \varphi_1) \right],
\end{equation}
where $V = [u_1^2 + 4u_0^2 + 4u_2^2 + 4v_1^2 v_2^2/(v_1^2 + v_2^2)]^{1/2}$, 
and $\theta_i, \varphi_i$ are the various properly normalized angular 
fields of the corresponding complex scalars.  The axionic coupling to 
quarks is thus
\begin{eqnarray}
&& (\partial_\mu a) {1 \over 2V} \left( {2v_1v_2 \over v_1^2+v_2^2} \right) 
\left[ {v_1 \over v_2} \bar{u} \gamma^\mu \gamma_5 u + {v_2 \over v_1} 
\bar{d} \gamma^\mu \gamma_5 d \right] \nonumber \\ && = {1 \over V} 
(\partial_\mu a) [\sin^2 \beta ~\bar{u} \gamma^\mu \gamma_5 u + 
\cos^2 \beta ~\bar{d} \gamma^\mu \gamma_5 d],
\end{eqnarray}
where $\tan \beta = v_1/v_2$, as in the DFSZ model.

The scale $m_2$ determines the axion scale as well as $m_N$ and $m_\Sigma$. 
The decay of the lightest $N^c$ generates a lepton asymmetry whereas the 
decay of the lightest $\Sigma$ generates a baryon asymmetry \cite{m08}.  
Below their common mass scale, both $L$ and $B$ are conserved additively, 
hence each asymmetry will be converted to a $B-L$ asymmetry through the 
interaction of the electroweak sphalerons \cite{dnn08}.  Whereas thermal 
equilibrium of leptons is affected by their known Yukawa couplings (flavor 
dependence), baryogenesis through $\Sigma$ decay may be more efficient 
because the $QQh$ and $u^cd^ch^c$ couplings may not necessarily be small. 
If kinematically allowed, $\tilde{h}$ and $\tilde{h}^c$ will be produced 
in abundance at the LHC.  Their decay into 2 quark jets may have a chance 
of being observed. 

The appearance of the exotic $h$ and $h^c$ superfields at the $M_{SUSY}$ 
scale, presumably of order TeV, would spoil the gauge-coupling unification 
of the MSSM at around $10^{16}$ GeV.  To remedy this situation, a very 
simple solution is to use four Higgs doublets instead of two.  This is 
easily accomplished by assigning them separately to the quark and lepton 
sectors: two to $d^c$ and $u^c$, and two to $e^c$ and $N^c$.  Now the two 
extra Higgs doublets combine with $h$ and $h^c$ to form complete multiplets 
of \underline{5} and \underline{5}$^*$ under $SU(5)$, thereby preserving 
the gauge-coupling unification of the MSSM.  It also means that the 
phenomenology of the supersymmetric Higgs sector becomes much richer. 

Since $(-)^L$ and $(-)^{3B}$ remain conserved, so is the usual $R$ parity 
of the MSSM.  The neutralino mass matrix is now $9 \times 9$ instead of 
$4 \times 4$ because of the two additional neutral higgsinos as well as 
$S_{2,1,0}$.  As shown in Eq.~(6), two of these fields combine to form 
a heavy Dirac fermion at the $m_0$ mass scale, leaving seven at the TeV 
scale.  One linear combination is then the axino, but its mixing with 
the other neutralinos is small, i.e. of order $v_{1,2}/V$.  The lightest 
among these seven particles is a candidate for the dark matter of the 
Universe, in addition to the axion. 

In conclusion, it has been pointed out in this note that the origin  
of the multiplicative conservation of lepton and baryon numbers may be 
the same as that of strong CP conservation.  They are all related to 
the complex singlet superfield $S_1$ whose vacuum expectation value 
determines the axion scale as well as the mass scale at which $L$ breaks 
to $(-)^L$ and $B$ breaks to $(-)^{3B}$.  There are then two sources for 
the baryon asymmetry of the Universe, thus relieving some of the 
tension inherent in the usual leptogenesis scenario \cite{dnn08}. 
Only one other mass scale appears in this scenario, i.e. $M_{SUSY}$, 
which explains why the electroweak breaking scale cannot be too far 
from it and bolsters the expectation that supersymmetry will be discovered 
at the LHC.  The implementation of $(-)^{3B}$ conservation predicts new 
exotic particles $h,h^c$ at the TeV scale, presumably together with two 
more Higgs doublets if gauge-coupling unification of the MSSM is to be 
maintained.  There are seven neutralinos (including the axino) at the 
TeV scale, the lightest of which is a dark-matter candidate, in addition 
to the axion.

This work was supported in part by the U.~S.~Department of Energy under Grant 
No.~DE-FG03-94ER40837.

\bibliographystyle{unsrt}

\begin{thebibliography}{99}

\bibitem{ev02} For a review, see for example S. R. Elliott and P. Vogel, 
Ann. Rev. Nucl. Part. Sci. {\bf 52}, 115 (2002).

\bibitem{m08} E. Ma, Phys. Lett. {\bf B661}, 273 (2008).

\bibitem{dnn08} For a review, see for example S. Davidson, E. Nardi, and 
Y. Nir, arXiv:0802.2962 [hep-ph].

\bibitem{cdg76} C. G. Callan, R. F. Dashen, and D. J. Gross, Phys. Lett. 
{\bf B63}, 334 (1976).

\bibitem{jr76} R. Jackiw and C. Rebbi, Phys. Rev. Lett. {\bf 37}, 172 (1976).

\bibitem{h99} P. G. Harris {\it et al.}, Phys. Rev. Lett. {\bf 82}, 904 (1999).

\bibitem{pq77} R. D. Peccei and H. R. Quinn, Phys. Rev. Lett. {\bf 38}, 1440 
(1977).

\bibitem{we78} S. Weinberg, Phys. Rev. Lett. {\bf 40}, 223 (1978).

\bibitem{wi78} F. Wilczek, Phys. Rev. Lett. {\bf 40}, 279 (1978).

\bibitem{rb00} For a review, see for example L. J. Rosenberg and K. A. 
van Bibber, Phys. Rep. {\bf 325}, 1 (2000).

\bibitem{r99} For a review, see for example G. G. Raffelt, Ann. Rev. Nucl. 
Part. Sci. {\bf 49}, 163 (1999).

\bibitem{dfs81} M. Dine, W. Fischler, and M. Srednicki, Phys. Lett. {\bf B104}, 
199 (1981).

\bibitem{z80} A. R. Zhitnitsky, Sov. J. Nucl. Phys. {\bf 31}, 260 (1980).

\bibitem{k79} J. E. Kim, Phys. Rev. Lett. {\bf 43}, 103 (1979).

\bibitem{svz80} M. A. Shifman, A. I. Vainshtein, and V. I. Zakharov, Nucl. 
Phys. {\bf B166}, 493 (1980).

\bibitem{dm00} D. A. Demir and E. Ma, Phys. Rev. {\bf D62}, 111901 (2000).

\bibitem{dms00} D. A. Demir, E. Ma, and U. Sarkar, J. Phys. {\bf G26}, L117 
(2000).

\bibitem{dm01} D. A. Demir and E. Ma, J. Phys. {\bf G27}, L87 (2001).

\bibitem{m07} E. Ma, Mod. Phys. Lett. {\bf A22}, 2721 (2007).

\bibitem{m01} E. Ma, Phys. Lett. {\bf B514}, 330 (2001).

\bibitem{m99} E. Ma, Mod. Phys. Lett. {\bf A14}, 1637 (1999).

\end{thebibliography}

\end{document}